\documentclass[12pt]{iopart}
\usepackage{graphicx}

\begin{document}

\title[R\'enyi-2 Quantum Correlations
of Two-Mode Gaussian Systems
]{R\'enyi-2 Quantum Correlations
of Two-Mode Gaussian Systems in a Thermal Environment }

\author{Aurelian Isar$^{1,2}$}

\address{$^1$National Institute of Physics and Nuclear Engineering,
P.O.Box MG-6, Bucharest-Magurele, Romania\\
$^2$Academy of Romanian Scientists, 54 Splaiul Independentei, Bucharest 050094, Romania}
\ead{isar@theory.nipne.ro}
\begin{abstract}

In the framework of the theory of open systems based on completely positive quantum dynamical semigroups, we give a description of continuous variable Gaussian R\'enyi-2 (GR2) quantum discord for a system consisting of two non-interacting non-resonant bosonic modes embedded in a thermal environment. In the case of both an entangled initial squeezed vacuum state and squeezed thermal state, GR2 quantum discord has non-negative values for all temperatures of the environment. We describe also the time evolution of GR2 classical correlations and GR2 quantum mutual information of the quantum system. All these quantities tend asymptotically to zero in the long time regime under the effect of the thermal bath.

\end{abstract}


\section{Introduction}

In recent years there is an increasing interest in using non-classical entangled states of continuous variable systems in applications of quantum information processing, communication and computation \cite{bra1,cer,ade1}. In this respect, Gaussian states, in particular two-mode Gaussian states, play a key role since they can be easily created and controlled experimentally. Due to the unavoidable interaction with the environment, in order to describe realistically quantum information processes it is necessary to take decoherence and dissipation into consideration. Decoherence and dynamics of quantum entanglement in continuous variable open systems have been intensively studied in the last years \cite{oli,ser3,pra,dod1,ser4,avd,ben1,mch,man,jan,aphysa,arus,aeur,arus1,paz2,arus2,ascri1}.

In quantum information theory an interesting family of additive entropies is represented by R\'{e}nyi-$\alpha$ entropies \cite{renyi}, defined by
\begin{equation}\label{ren}
{\cal S}_\alpha(\rho) = (1-\alpha)^{-1} \ln {\rm Tr} (\rho^\alpha).
\end{equation}
In the limit $\alpha \rightarrow 1$ they reduce to the von Neumann entropy ${\cal S}(\rho) = - {\rm Tr}(\rho \ln \rho),$ which in quantum information theory quantifies the degree of information contained in a quantum state $\rho$, and in the case $\alpha=2$ we obtain ${\cal S}_2(\rho) = - \ln {\rm Tr} (\rho^2),$ which is a quantity directly related to the purity of the state. Von Neumann entropy satisfies the strong subadditivity inequality which is a key requirement for quantum information theory, implying in particular that the mutual information, which measures the total correlations between two subsystems in a bipartite state, is always non-negative. R\'{e}nyi entropies are powerful quantities for studying quantum correlations in multipartite states \cite{man1,man2}. In general R\'{e}nyi-$\alpha$ entropies for $\alpha \neq 1$ are not subadditive. In Ref. \cite{ade0} it was demonstrated that R\'{e}nyi-$2$ entropy provides a natural measure of information for any multimode Gaussian state of quantum harmonic systems. It was proven that for all Gaussian states this entropy satisfies the strong subadditivity inequality, which made possible to define measures of Gaussian R\'{e}nyi-$2$ (GR2) entanglement, total, classical, and discord-like quantum correlations based on this entropy. In this sense one could regard R\'{e}nyi-$2$ entropy as a specially meaningful choice to develop a Gaussian theory of quantum information and correlations \cite{ade0}.

In this paper we study, in the framework of the theory of open systems based on completely positive quantum dynamical semigroups, the dynamics of continuous variable GR2 quantum discord of a subsystem consisting of two bosonic modes (harmonic oscillators) interacting with a common thermal environment. In Sec. 2 we give the general solution of the evolution equation for the covariance matrix, i.e. we derive the variances and covariances of coordinates and momenta. We are interested in discussing the correlation effect of the environment, therefore we assume that the two modes are independent, i.e. they do not interact directly. The initial state of the subsystem is taken of Gaussian form and the evolution under the quantum dynamical semigroup assures the preservation in time of the Gaussian form of the state. In Sec. 3 we show that in the case of both an entangled initial squeezed vacuum state and squeezed thermal state, GR2 quantum discord has non-negative values in time for all values of the temperature of the environment. We describe also the time evolution of GR2 classical correlations and GR2 quantum mutual information, which measures the total amount of GR2 correlations of the quantum system.. All these quantities tend asymptotically to zero in time under the effect of the thermal bath. A summary is given in Sec. 4.

\section{Equations of motion for two modes interacting with an environment}

We study the dynamics of a subsystem composed of two non-interacting bosonic modes in weak interaction with a thermal environment. In the axiomatic formalism based on completely positive quantum dynamical semigroups, the Markovian irreversible time evolution of an open system is described by the Kossakowski-Lindblad master equation \cite{rev,san}.
We are interested in the set of Gaussian states, therefore we introduce such quantum dynamical semigroups that preserve this set during time evolution of the system. The Hamiltonian of the two uncoupled non-resonant harmonic oscillators of identical mass $m$ and frequencies $\omega_1$ and $\omega_2$ is
\begin{eqnarray} H={1\over 2m}(p_x^2+p_y^2)+\frac{m}{2}(\omega_1^2 x^2+\omega_2^2 y^2),\end{eqnarray} where $x,y$ are the coordinates and $p_x,p_y$ are the momenta of the oscillators.

The equations of motion for the quantum correlations of the canonical observables $x,y$ and $p_x,p_y$ are the following ($\rm T$ denotes the transposed matrix) \cite{san}:
\begin{eqnarray}{d \sigma(t)\over
dt} = Y \sigma(t) + \sigma(t) Y^{\rm T}+2 D,\label{vareq}\end{eqnarray} where
\begin{eqnarray} Y=\left(\matrix{ -\lambda&1/m&0 &0\cr -m\omega_1^2&-\lambda&0&
0\cr 0&0&-\lambda&1/m \cr 0&0&-m\omega_2^2&-\lambda}\right), \nonumber \\
D=\left(\matrix{
D_{xx}& D_{xp_x} &D_{xy}& D_{xp_y} \cr D_{xp_x}&D_{p_x p_x}&
D_{yp_x}&D_{p_x p_y} \cr D_{xy}& D_{y p_x}&D_{yy}& D_{y p_y}
\cr D_{xp_y} &D_{p_x p_y}& D_{yp_y} &D_{p_y p_y}} \right),\end{eqnarray}
and the diffusion coefficients $D_{xx}, D_{xp_x},$... and the dissipation constant $\lambda$ are real quantities. We introduced the following $4\times 4$ bimodal covariance matrix:
\begin{eqnarray}\sigma(t)=\left(\matrix{\sigma_{xx}(t)&\sigma_{xp_x}(t) &\sigma_{xy}(t)&
\sigma_{xp_y}(t)\cr \sigma_{xp_x}(t)&\sigma_{p_xp_x}(t)&\sigma_{yp_x}(t)
&\sigma_{p_xp_y}(t)\cr \sigma_{xy}(t)&\sigma_{yp_x}(t)&\sigma_{yy}(t)
&\sigma_{yp_y}(t)\cr \sigma_{xp_y}(t)&\sigma_{p_xp_y}(t)&\sigma_{yp_y}(t)
&\sigma_{p_yp_y}(t)}\right)=\left(\begin{array}{cc}A&C\\
C^{\rm T}&B \end{array}\right),\label{covar}
\end{eqnarray}
where $A$, $B$ and $C$ are $2\times 2$ Hermitian matrices. $A$ and $B$ denote the symmetric covariance matrices for the individual reduced one-mode states, while the matrix $C$ contains the cross-correlations between modes.
The elements of the covariance matrix are defined as $\sigma_{ij}={\rm Tr}[\rho\{R_iR_j+R_jR_i\}]/2, i,j=1,..,4,$ with ${\bf R}=\{x,p_x,y,p_y\},$ which up to local displacements fully characterize any Gaussian state of a bipartite system \cite{aijqi}.
All the measures defined below are invariant under local unitaries, so we will assume our states to have zero first moments, $\langle {\bf R}\rangle = {\bf 0}$, without loss of generality.

The time-dependent solution of Eq. (\ref{vareq}) is given by \cite{san}
\begin{eqnarray}\sigma(t)= M(t)[\sigma(0)-\sigma(\infty)] M^{\rm T}(t)+\sigma(\infty),\label{covart}\end{eqnarray} where the matrix $M(t)=\exp(Yt)$ has to fulfill the condition $\lim_{t\to\infty} M(t) = 0.$ The values at infinity are obtained from the equation \begin{eqnarray}
Y\sigma(\infty)+\sigma(\infty) Y^{\rm T}=-2 D.\label{covarinf}\end{eqnarray}

\section{Dynamics of GR2 correlations}

\subsection{Time evolution of GR2 quantum discord}

The GR2 mutual information ${\cal I}_2$ for an arbitrary bipartite Gaussian state $\rho_{AB}$ is defined by
\begin{eqnarray}
{\cal I}_2(\rho_{A:B})&=& {\cal S}_2(\rho_A) + {\cal S}_2(\rho_B) - {\cal S}_2(\rho_{AB}),\label{mutinf}\end{eqnarray}
where $\rho_A$ and $\rho_B$ are the two marginals of $\rho_{AB}$. It was shown \cite{ade0} that ${\cal I}_2(\rho_{A:B})\ge 0$ and it measures the total quadrature correlations of  $\rho_{AB}$.

One can also define a GR2 measure of one-way classical correlations ${\cal J}_2(\rho_{A|B})$  \cite{ade0,hen}, as the maximum decrease in the R\'{e}nyi-$2$ entropy of subsystem $A$, given a Gaussian measurement has been performed on subsystem $B$, where the maximization is over all Gaussian measurements, which map Gaussian states into Gaussian states \cite{par,ade}.

Following \cite{oll} and the analysis of Gaussian quantum discord using von Neumann entropy \cite{par,ade}, in \cite{ade0} it was defined a Gaussian non-negative measure of quantumness of correlations based on R\'{e}nyi-$2$ entropy, namely GR2 discord, as the difference between mutual information (\ref{mutinf}) and classical correlations:
\begin{equation}
{\cal D}_2(\rho_{A|B})= {\cal I}_2(\rho_{A:B})- {\cal J}_2(\rho_{A|B}).\label{disc}
\end{equation}
For a general two-mode Gaussian state $\rho_{AB}$, with $A$ and $B$ single modes, closed formulae have been obtained for GR2 classical correlations and GR2 discord \cite{ade0}. For pure bipartite Gaussian states $\rho_{AB}$ the following equalities hold:
$\frac12 {\cal I}_2(\rho_{A:B})={\cal J}_2(\rho_{A|B})={\cal D}_2(\rho_{A|B})={\cal S}_2(\rho_A).$

The covariance matrix $\sigma$ (\ref{covar}) of any two-mode Gaussian state can be transformed, by means of local unitary operations, into a standard form of the type \cite{ade2}
\begin{equation}
\label{covar1}
\sigma=\left(\begin{array}{cc}
A&C\\
C^{\rm T}&B
\end{array}\right) = \frac 12 \left(\begin{array}{cccc}
a&0&c_{+}&0\\
0&a&0&c_{-}\\
c_{+}&0&b&0\\
0&c_{-}&0&b
\end{array}\right),\label{stand}
\end{equation}
where $a,b\geq 1$, $\left[\left(a^2-1\right) \left(b^2-1\right)-a b \left(c_-^2+c_+^2\right)-2 c_- c_++c_-^2 c_+^2\right] \geq 0$, and we can set $c_+ \ge |c_-|$ without losing generality. These conditions ensure that the uncertainty relation  $\sigma \geq i {\bf \omega}^{\oplus 2}/16$ is satisfied (${\bf \omega} = {{\ 0 \ \ 1} \choose {-1 \ 0}}$ is the symplectic matrix), which is a requirement for the covariance matrix $\sigma$ to be associated with a physical Gaussian state in a two-mode infinite-dimensional Hilbert space \cite{ade2}.
For pure Gaussian states, $b=a$, $c_+=-c_-=\sqrt{a^2-1}$.

For generally mixed two-mode Gaussian states $\rho_{AB}$, the R\'{e}nyi-$2$ measure of one-way quantum discord (\ref{disc}) has the following expression if the covariance matrix $\sigma$ is in standard form (\ref{stand}) \cite{ade0}:
\begin{eqnarray}
{\cal D}_2(\rho_{A|B}) &=& \ln b - \frac12 \ln (\det C) + \frac12 \ln \varepsilon_2, \label{d2}
\end{eqnarray}
where
\begin{equation}\label{det}
\varepsilon_2=  \\
\left\{
\begin{array}{c}
 a \left(a-\frac{c_+^2}{b}\right), {\rm if} \left(a b^2 c_-^2-c_+^2 \left(a+b c_-^2\right)\right) \left(a b^2 c_+^2-c_-^2 \left(a+b c_+^2\right)\right)<0; \\
 \frac{2 \left|c_- c_+\right| \sqrt{\left(a \left(b^2-1\right)-b c_-^2\right) \left(a \left(b^2-1\right)-b c_+^2\right)}+\left(a \left(b^2-1\right)-b c_-^2\right) \left(a \left(b^2-1\right)-b c_+^2\right)+c_-^2 c_+^2}{\left(b^2-1\right)^2}, {\rm otherwise.}
\end{array}
\right.
\end{equation}
These formulae, written explicitly for standard form covariance matrices, can be reobtained in a locally invariant form by expressing them in terms of the four local symplectic invariants of a generic two-mode Gaussian state \cite{ser5}, $I_1= 4 \det A$, $I_2= 4 \det B$, $I_3= 4 \det C$, $I_4= 16 \det\sigma$ \cite{aosid}. This can be realized by inverting the relations $I_1=a^2, I_2=b^2, I_3=c_+ c_-, I_4=(a b -c_+)(a b - c_-)$ \cite{ade2}. The obtained expressions are then valid for two-mode covariance matrices in any symplectic basis, beyond the standard form.

We assume that the initial Gaussian state is a two-mode squeezed thermal state, with the covariance matrix of the form \cite{mar,ascri2}
\begin{eqnarray}\sigma_{s}(0)=\frac{1}{2}\left(\matrix{a_s&0&c_s&0\cr
0&a_s&0&-c_s\cr
c_s&0&b_s&0\cr
0&-c_s&0&b_s}\right),\label{ini1} \end{eqnarray}
with the matrix elements given by
\begin{eqnarray}a_s=n_1 \cosh^2 r + n_2 \sinh^2 r + \frac{1}{2} \cosh 2r,\\
b_s=n_1 \sinh^2 r + n_2 \cosh^2 r + \frac{1}{2} \cosh 2r,\\
c_s=\frac{1}{2}(n_1 + n_2 + 1) \sinh 2r,\label{ini2}
\end{eqnarray}
where $n_1,n_2$ are the average number of thermal photons associated with the two modes and $r$ denotes the squeezing parameter.
In the particular case $n_1=0$ and $n_2=0$, (\ref{ini1}) becomes the covariance matrix of the two-mode squeezed vacuum state \cite{aosid1}. A two-mode squeezed thermal state is entangled when the squeezing parameter $r$ satisfies the inequality $r>r_s$ \cite{mar},
where \begin{eqnarray} \cosh^2 r_s=\frac{(n_1+1)(n_2+1)}{ n_1+n_2+1}. \end{eqnarray}

We suppose that the asymptotic state of the considered open system is a Gibbs state corresponding to two independent bosonic modes in thermal equilibrium at temperature $T$ \cite{ascri}. Then the quantum diffusion coefficients have the following form (we put $\hbar=1$) \cite{rev}:
\begin{eqnarray}m\omega_1 D_{xx}=\frac{D_{p_xp_x}}{m\omega_1}=\frac{\lambda}{2}\coth\frac{\omega_1}{2kT},\nonumber\\
m\omega_2 D_{yy}=\frac{D_{p_yp_y}}{m\omega_2}=\frac{\lambda}{2}\coth\frac{\omega_2}{2kT},\label{envcoe}\\
D_{xp_x}=D_{yp_y}=D_{xy}=D_{p_xp_y}=D_{xp_y}=D_{yp_x}=0.\nonumber\end{eqnarray}

The evolution of the GR2 quantum discord is illustrated in Figs. 1 and 2, where we represent the dependence of ${\cal D}_2(\rho_{A|B})$ on time $t$ and temperature $T$ for an initial entangled Gaussian state with the covariance matrix given by Eq. (\ref{ini1}), taken of the form of a two-mode squeezed vacuum state, respectively squeezed thermal state, and for such values of the parameters that satisfy for all times the first condition in formula (\ref{det}). The GR2 discord has positive values for all finite times and in the limit of infinite time it tends asymptotically to zero, corresponding to the thermal product (separable) state, with no correlations at all. We also notice that the decay of GR2 discord is stronger when the temperature $T$ and dissipation constant $\lambda$ are increasing.

\begin{figure}
{
\includegraphics{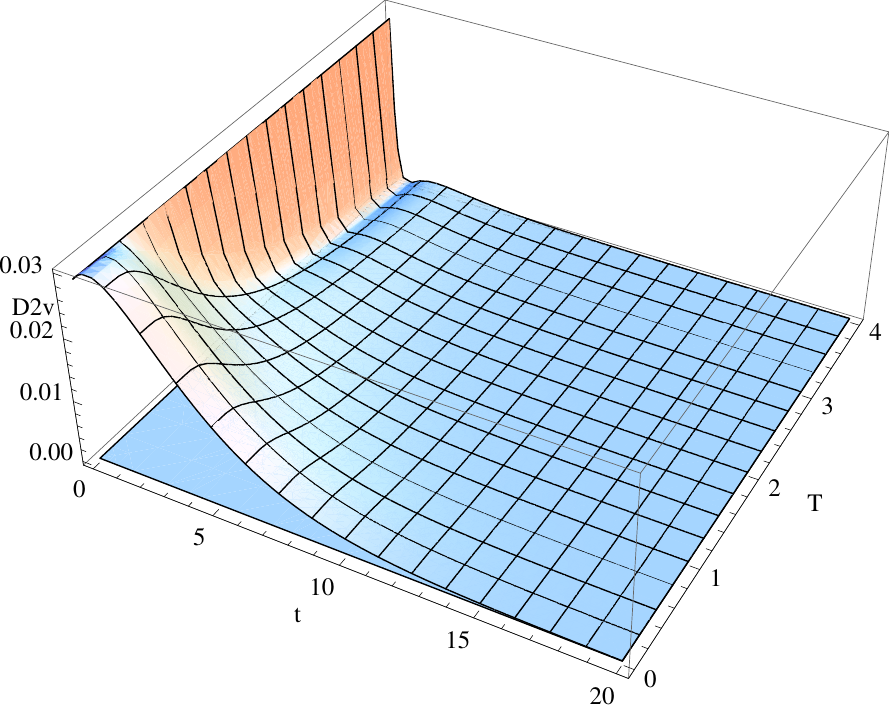}
}
\caption{GR2 quantum discord ${\cal D}_{2v}(\rho_{A|B})$ versus time $t$ and temperature $T$ for an entangled initial squeezed vacuum state with squeezing parameter $r=0.23$, $n_1=0, n_2=0$ and $\lambda=0.1, \omega_1=1, \omega_2=2.$ We take $m=\hbar=k=1.$
}
\label{fig:1}
\end{figure}

\begin{figure}
{
\includegraphics{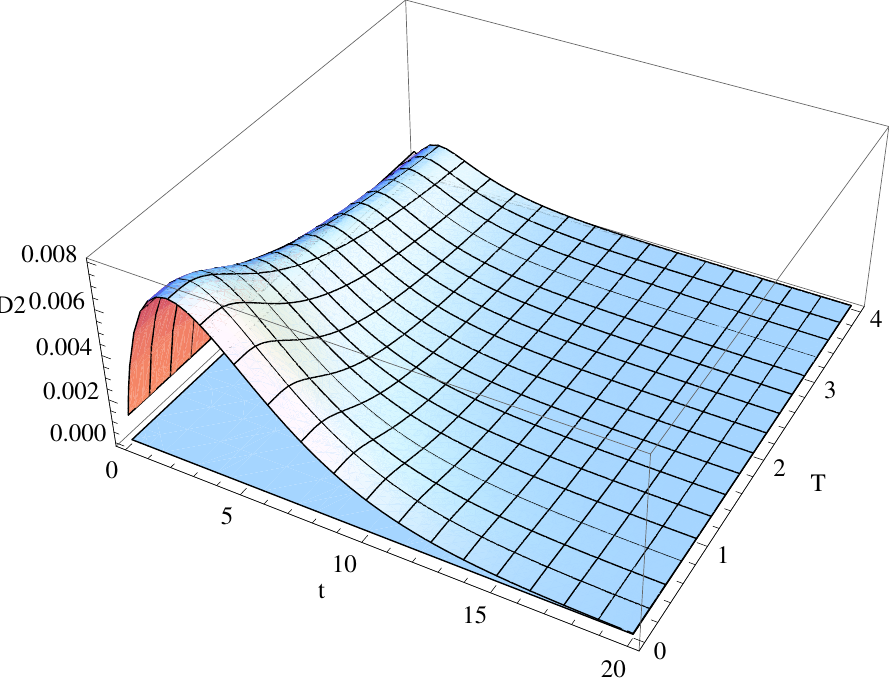}
}
\caption{GR2 quantum discord ${\cal D}_2(\rho_{A|B})$ versus time $t$ and temperature $T$ for an entangled initial non-symmetric squeezed thermal state with squeezing parameter $r=0.1$, $n_1=1, n_2=0$ and $\lambda=0.1, \omega_1=1, \omega_2=2.$ We take $m=\hbar=k=1.$
}
\label{fig:2}
\end{figure}

\subsection{GR2 classical correlations and quantum mutual information}

For a mixed two-mode Gaussian state $\rho_{AB}$, GR2 measure of one-way classical correlations ${\cal J}_2(\rho_{A|B})$ has the following expression if the covariance matrix $\sigma$ is in standard form (\ref{stand}) \cite{ade0}:
\begin{eqnarray}
{\cal J}_2(\rho_{A|B}) &=& \ln a -    \frac12 \ln \varepsilon_2, \label{c2}
\end{eqnarray}
where $\varepsilon_2$ is given by Eq. (\ref{det}), while the expression of the GR2 quantum mutual information, which measures the total correlations, is given by
\begin{eqnarray}
{\cal I}_2(\rho_{A:B}) = \ln a+ \ln b - \frac12 \ln (\det C).
\end{eqnarray}

\begin{figure}
{
\includegraphics{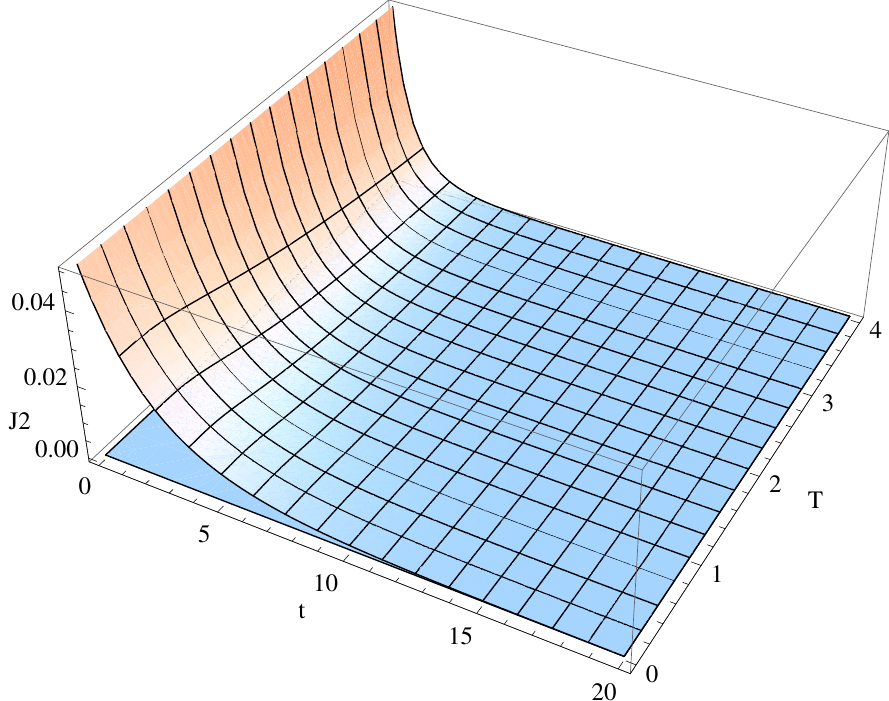}
}
\caption{GR2 classical correlations ${\cal J}_2(\rho_{A|B})$ versus time $t$ and temperature $T$ for an entangled initial non-symmetric squeezed thermal state with squeezing parameter $r=0.1$, $n_1=1, n_2=0$ and $\lambda=0.1, \omega_1=1, \omega_2=2.$ We take $m=\hbar=k=1.$
}
\label{fig:3}
\end{figure}

\begin{figure}
{
\includegraphics{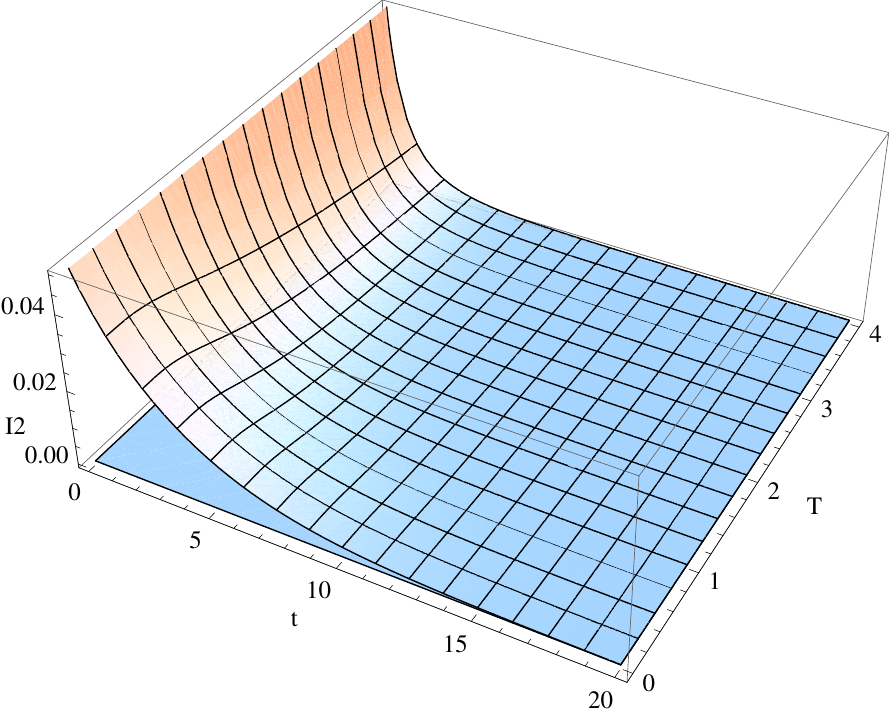}
}
\caption{GR2 quantum mutual information ${\cal I}_2(\rho_{A:B})$ versus time $t$ and temperature $T$ for an entangled initial non-symmetric squeezed thermal state with squeezing parameter $r=0.1$, $n_1=1, n_2=0$ and $\lambda=0.1, \omega_1=1, \omega_2=2.$ We take $m=\hbar=k=1.$
}
\label{fig:4}
\end{figure}

In Figs. 3 and 4 we illustrate the evolution of classical correlations ${\cal J}_2(\rho_{A|B})$ and, respectively, quantum mutual information ${\cal I}_2(\rho_{A:B})$ as functions of time $t$ and temperature $T$ for an entangled initial Gaussian state, taken of the form of a two-mode squeezed thermal state (\ref{ini1}), and for such values of the parameters that satisfy for all times the first condition in formula (\ref{det}). Both these quantities manifest a similar qualitative behaviour similar: they have positive values for all finite times and in the limit of infinite time they tend asymptotically to zero, corresponding to the thermal product (separable) state, with no correlations at all. One can also see that the classical correlations and quantum mutual information decrease with increasing the temperature of the thermal bath and the dissipation coefficient. For comparison, GR2 mutual information, classical correlations and discord are represented also on the same graphic in Fig. 5.  In the considered case the value of GR2 classical correlations is larger than that of quantum correlations, represented by the GR2 quantum discord.

\begin{figure}
{
\includegraphics{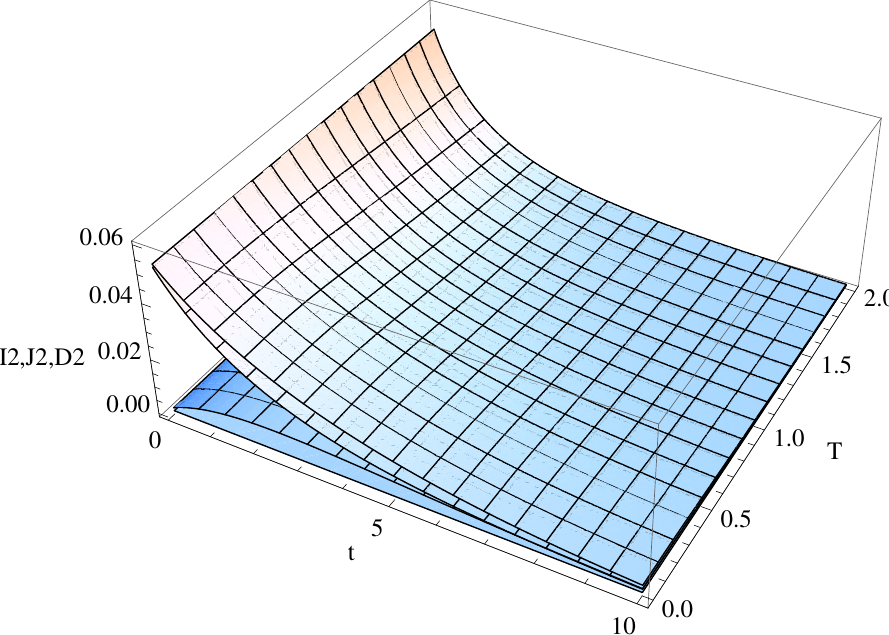}
}
\caption{GR2 quantum mutual information, classical correlations and quantum discord versus time $t$ and temperature $T$ for an entangled initial non-symmetric squeezed thermal state with squeezing parameter $r=3$, $n_1=3, n_2=1$ and $\lambda=0.1, \omega_1=1, \omega_2=2.$ We take $m=\hbar=k=1.$
}
\label{fig:5}
\end{figure}

\section{Summary}

We investigated the Markovian dynamics of Gaussian R\'enyi-2 quantum correlations for a subsystem composed of two non-interacting bosonic modes embedded in a thermal bath. We have analyzed the influence of the environment on the dynamics of GR2 quantum discord, classical correlations and quantum mutual information, which measures the total GR2 correlations of the quantum system, for initial squeezed vacuum states and non-symmetric squeezed thermal states, for the case when the asymptotic state of the considered open system is a Gibbs state corresponding to two independent quantum harmonic oscillators in thermal equilibrium. The dynamics of these quantities strongly depend on the initial states and the parameters characterizing the environment (dissipation coefficient and temperature). Their values decrease asymptotically in time with increasing the temperature and dissipation.

The study of time evolution of these measures may contribute to understanding the quantification of correlations defined on the base of the R\'{e}nyi-$2$ entropy, which could represent a special tool in developing a Gaussian theory of quantum information and correlations.

\ack

The author acknowledges the financial support received from the Romanian Ministry of Education and Research, through the Projects CNCS-UEFISCDI PN-II-ID-PCE-2011-3-0083 and PN 09 37 01 02/2010.

\end{document}